\documentclass[letterpaper, 10 pt, conference]{ieeeconf}  % Comment this line out if you need a4paper

\IEEEoverridecommandlockouts                              % This command is only needed if 
\usepackage{cite}
\usepackage{amsmath,amssymb,amsfonts}
\usepackage{graphicx}
\usepackage{subcaption}
\usepackage{textcomp}
\usepackage{color}
\usepackage{bbm}
\usepackage{balance}
\usepackage{tikz}
\usetikzlibrary{automata, positioning, arrows.meta, calc, fit, backgrounds, shadows, decorations.pathmorphing}
\usepackage[american]{circuitikz}

\usepackage{booktabs}
\usepackage{tabularx}
\usepackage{multirow} % Required for merging rows
\usepackage{xcolor}
\newtheorem{thm}{Theorem}

\newtheorem{fact}{Fact}
\newtheorem{assumption}{Assumption}

\newtheorem{defi}{Definition}
\newtheorem{prop}{Proposition}

\newtheorem{exx}{Example}
\newtheorem{remm}{Remark}

\usepackage{bbold}

\usepackage{xcolor}
\usepackage[font=footnotesize, labelfont=footnotesize]{caption}
\newenvironment{assume}{\begin{assumption}\rm }{\hfill \hspace*{1pt} \hfill $\lrcorner$ \end{assumption}}

\newenvironment{proposition}{\begin{prop}\rm }{\hfill \hspace*{1pt} \hfill $\lrcorner$ \end{prop}}

\newcommand\real{\ensuremath{{\mathbb R}}}

\newcommand{\calV}{\mathcal{V}}

\definecolor{Leak}{HTML}{AAAAAA}
\definecolor{Sodium}{HTML}{12BB2C}
\definecolor{Potassium}{HTML}{3960C9}

\title{\LARGE \bf
Spiking neuromodulation: a linear control perspective
}

\author{Fulvio Forni$^{1}$ and Rodolphe Sepulchre$^{1,2}$% <-this % stops a space
\thanks{$^{1}$ University of Cambridge, Department of Engineering, Trumpington Street, CB2 1PZ Cambridge, United Kingdom.} %
\thanks{$^{2}$ KU Leuven, Department of Electrical Engineering (STADIUS), KasteelPark Arenberg, 10, B-3001 Leuven, Belgium.}%
\thanks{Emails: f.forni@eng.cam.ac.uk, rodolphe.sepulchre@kuleuven.be.}%
\thanks{The research leading to these results has received funding from the European Research Council under the Advanced
ERC Grant Agreement SpikyControl n.101054323.}% <-this % stops a space
}

\begin{document}

\maketitle
\thispagestyle{empty}
\pagestyle{empty}

%%%%%%%%%%%%%%%%%%%%%%%%%%%%%%%%%%%%%%%%%%%%%%%%%%%%%%%%%%%%%%%%%%%%%%%%%%%%%%%%
\begin{abstract}

Neuronal excitability is a nonlinear phenomenon,
and the complexity of spiking experimental traces sometimes looks defeating.
The present paper takes the view point that excitability is a neuromodulated
property, and aims to demonstrate that the neuromodulation of spiking is amenable to
the classical linear analysis tools of control theory.
We illustrate the value of linear feedback analysis as an
alternative to bifurcation analysis in the simple yet classical
example of spiking modulation by  the A-type potassium current.

\end{abstract}

%%%%%%%%%%%%%%%%%%%%%%%%%%%%%%%%%%%%%%%%%%%%%%%%%%%%%%%%%%%%%%%%%%%%%%%%%%%%%%%%
\section{INTRODUCTION}

The concept of neuromodulation  has become central
to our understanding of animal brains \cite{marder2014}. Neuronal circuits are continuously
modulated by neuromodulatory mechanisms spanning a whole range of temporal
and spatial scales, and those regulatory mechanisms are essential to any form of
animal behavior. This evolution of neuroscience is an invitation to control
scientists to regard neuronal excitability and its regulation by neuromodulators 
as a control problem\cite{sepulchre2019Control}. The current development of neuromorphic control
further exacerbates the need for a control theory of neuronal excitability  \cite{sepulchre2022}.

A longstanding methodological obstacle to this program
is the apparent mismatch between  neuronal excitability, when regarded as an unstable nonlinear phenomenon,
and regulation theory, concerned with  shaping the sensivity of a stable equilbrium.
For the last sixty years, the mathematical modelling of excitability has been
 a branch of dynamical system theory, see e.g.  \cite{izhikevich2007dynamical} for an excellent introduction. The analysis tools
 include  phase portrait analysis, bifurcation analysis, and singularity theory.
 Control theory plays no role in that theory.

The purpose of the present paper is to explore the potential of classical {\it linear control theory} to analyse and design
excitable systems. The novelty is not in the use of linearized models, which is most classical \cite{mauro1970subthreshold,sabah1969subthreshold,chandler1962theoretical}, \cite[Ch.10]{koch1999biophysics}. The novelty is not either in the neuromodulation perspective, which has been the topic of
a long line of work by the second author and colleagues \cite{drion2012,franci2012,franci2018,sepulchre2019Control}.
The novelty is rather in adopting a strictly control-theoretic perspective: we trade the language of dynamical system theory for
the most classical language of control theory. 

Our starting point is the mixed-feedback representation of an excitable neuron \cite{sepulchre2019}, and the linearization 
of this nonlinear control system around its equilibria. We extract two key properties of this feedback system: a right-half
plane zero of the controller, and the classical stability margin of the control system, measured as 
the distance of the Nyquist curve to the critical point in the complex plane. We show that those two quantities
can be defined in arbitrary conductance-based models, and they are structural rather than parameter dependent. 
Then we show how neuromodulation can regulate those two quantities and how this translates into the regulation
of neuronal spiking. We argue that the pole zero configuation
of the controller determines the {\it type} of excitability of the neuron, and that the stability margin
is an excellent indicator of the sensitivity of the neuron, that is, its propensity to spike. 

For the sake of illustration, we illustrate the proposed methodology on two distinct paradigmatic models: Fitzhugh-Nagumo
model \cite{fitzhugh1961impulses} and Connor-Stevens model \cite{connor1977neural}. 
The former is arguably the simplest mathematical model of excitability. We use it to introduce
the main ideas of the paper and to contrast them with the bifurcation approach of neurodynamics. 
The latter is used to apply the proposed analysis to the  conductance-based model of neurophysiology.
Historically, the model of Connor and Stevens is one of the most popular extensions of the seminal model of Hodgkin and Huxley: it includes one regulatory current (the so-called A-type current) in addition to the two ionic currents of Hodgkin-Huxley model that characterize the essential mixed-feedback mechanism of excitability. 
We use this model to illustrate that its regulation is primarily achieved by the two proposed features.

\section{A motivating example}
\label{sec:FN}

\subsection{FitzHugh-Nagumo model}
\label{sec:FN_hopf}

The model of FitzHugh and Nagumo
\cite{fitzhugh1961impulses,Nagumo1962pulsetransmission,sherwood2014fitzhughnagumo} is arguably the simplest mathematical model of excitability, amenable to phase portrait analysis:
\begin{equation}
\label{eq:fhn_tau}
\begin{aligned}
C\dot{v}&= \underbrace{v - \frac{v^3}{3} - w}_{-i_{\mathrm{int}}} + i_{\mathrm{ext}}, \\
\tau \dot{w} &= v + a - b w.
\end{aligned}
\end{equation}
$v$ is the membrane voltage, driven by internal and external currents, $i_\mathrm{int}$ and $i_{\mathrm{ext}}$. 
In phase-portrait analysis, the offset and slope parameters $a$ and $b$ regulate the intersection of the $w$-nullcline, $w = \frac{v+a}{b},$ 
with the  $v$-nullcline, 
$w = v-\frac{v^3}{3}$. 
Typically, $0 < b < 1$, $1-\frac{2b}{3} < a < 1$ and $\tau > b$.
The capacitance $C$ is typically normalized to $C=1$. 

Linearized analysis of FitzHugh Nagumo model is classical. For every constant $i_\mathrm{ext}$, there is a corresponding 
equilibrium $(v_0,w_0)$. The equilibrium becomes unstable when one of the two coefficients of the characteristic polynomial
\begin{equation}
s^2 + 
\underbrace{\left(\frac{b}{\tau} - (1 - v_0^2) \right)}_{c_1(v_0)} s + 
\underbrace{\frac{1 - b(1 - v_0^2)}{\tau}}_{c_0(v_0)}
\end{equation}
becomes negative.  Instability occurs for $c_1(v_0) < 0$,
that is, 
$
v_0^2 < 1 - \frac{b}{\tau}$.
At the threshold  
$|v_0| = \sqrt{1 - \frac{b}{\tau}}$ the two poles are purely imaginary if 
$c_0(v_0) > 0$, that is, if
$v_0^2 > 1 - \frac{1}{b}$.
This requires $\tau > b^2$, which is always guaranteed for the typical ranges
$0 < b < 1$ and $\tau > b$. This linear analysis identifies the Hopf bifurcation of the model.
 
\begin{figure}[htbp]
    \centering
    % Top Row: Poles & Spiking
    \vspace{2mm}
    \begin{subfigure}[b]{0.45\linewidth}
        \centering
        \includegraphics[width=\linewidth]{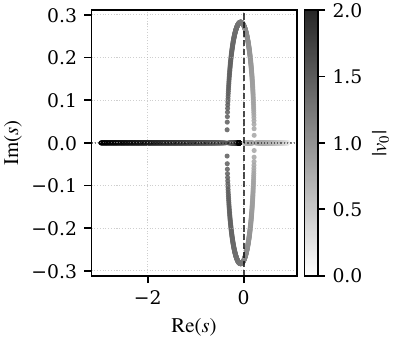}
        \caption{}
        \label{fig:FN_closedloopHopf}
    \end{subfigure}%
        % \hfill
    \begin{subfigure}[b]{0.45\linewidth}
        \centering\vspace{2mm}
        \includegraphics[width=\linewidth]{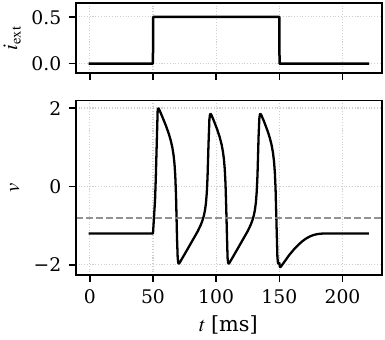}
        \caption{}
        \label{fig:FN_voltagetrajectories}
    \end{subfigure} 

    \vspace{0.25cm}

    % Bottom Row: Filtered Responses (Slow & Fast)
    \begin{subfigure}[b]{0.45\linewidth}
        \centering
        \includegraphics[width=\linewidth]{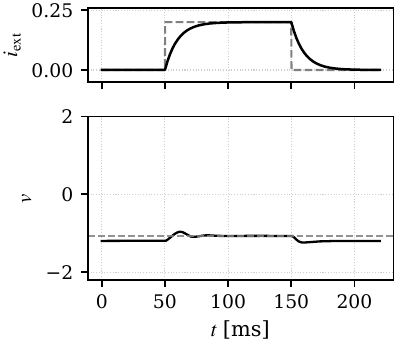}
        \caption{}
        \label{fig:FN_filtered_slow}
    \end{subfigure}%
    % \hfill
    \begin{subfigure}[b]{0.45\linewidth}
        \centering
        \includegraphics[width=\linewidth]{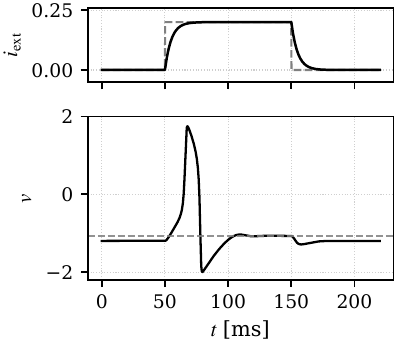}
        \caption{}
        \label{fig:FN_filtered_fast}
    \end{subfigure}

    \caption{FitzHugh--Nagumo model: (a) poles of $T_{\mathrm{FN}}(s)$ at voltages $0 \leq |v_0| \leq 2$, (b) behavior for $i_{\mathrm{ext}}=0.5$, where the dashed line represents the equilibrium $v_0 \simeq -0.805$ past Hopf bifurcation, and (c)–(d) behavior for filtered $i_\mathrm{ext}(t) \leq 0.2 < i_\mathrm{ext}^*$ with time constants $\tau_f = 10$ and $\tau_f = 5$.}
    \label{fig:FN_2x2_grid}
\end{figure}

An illustration
 is provided in Fig. \ref{fig:FN_2x2_grid},
 for standard values $a=0.7$, $b=0.8$ and $\tau = 12.5$. 
The Hopf bifurcation occurs at $v_0^* \simeq -0.97$, which corresponds to the external current $i_{\mathrm{ext}}^* \simeq 0.33$. This provides a raw form of spiking threshold. Smaller dynamic currents can trigger a spike if they generate a large enough overshoot in the voltage response.

\subsection{FitzHugh-Nagumo as feedback interconnection}
\label{sec:FN_nyquist}

This paper regards FitzHugh-Nagumo model as the feedback control system shown in Fig. \ref{fig:neural_block_diagram}:   the ``plant'' is the passive membrane capacitance
\begin{equation}
G(s) := \frac{1}{Cs}
\end{equation}
and the linearized controller is 
\begin{subequations}
\begin{equation}
\delta i_{\mathrm{int}}(s) 
= 
K(v_0,s) \delta v(s) 
\end{equation}
for 
\begin{equation}
K(v_0,s) :=  -(1-v_0^2) + \frac{1}{\tau s + b}. 
\label{eq:G_ionic_FN}
\end{equation}
The linearized transfer function from $i_\mathrm{ext}$ to $v$ of the closed-loop system reads
\begin{equation}
\begin{aligned}
\frac{1}{s\!+\!K(v_0,s)} = \frac{\tau s + b}{\tau s^2 \!+\! \big(b - \tau(1 \!-\! v_0^2)\big) s \!+\! 1 -\! b(1 - v_0^2)},
\end{aligned} 
\label{eq:FN_closedloop_TF}
\end{equation}
\end{subequations}
which, based on Fig. \ref{fig:neural_block_diagram},
corresponds to
\begin{subequations}
\begin{equation}
\begin{aligned}
T(v_0,s)
& := \frac{G(s)}{1 + G(s)K(v_0,s)} = \frac{\frac{1}{Cs}}{1 + \frac{K(v_0,s)}{Cs}}.
\end{aligned}
\end{equation}
Stability can thus be analyzed via the Nyquist diagram of the return ratio
\begin{equation}
\label{eq:return_ratio}
L(v_0,s) := G(s)K(v_0,s) = \frac{K(v_0,s)}{Cs}.
\end{equation}
\end{subequations}
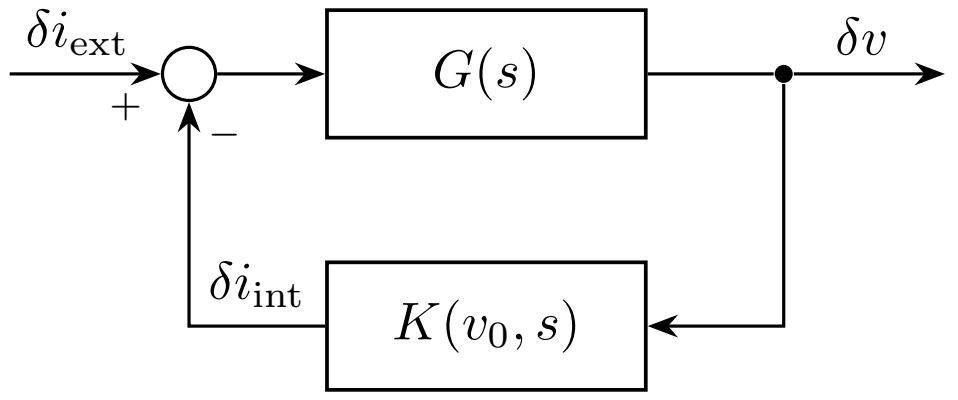
\begin{figure}[htbp]
  \centering
  \begin{tikzpicture}[scale=0.75,
      >=Stealth,
      node distance=0.8cm and 1.2cm,
      auto,
      block/.style={
        draw, 
        rectangle, 
        minimum height=2.4em, 
        minimum width=6.0em, % Standardized block width for alignment
        align=center,
        line width=0.7pt % Restored standard line width
      },
      sum/.style={
        draw, 
        circle, 
        minimum size=1em, 
        inner sep=0pt,
        line width=0.6pt
      }
    ]

    % Central Blocks
    \node [block] (membrane) {$G(s)$};
    \node [block, below=0.8cm of membrane] (ionic) {$K(v_0,s)$};

    % Equidistant Junctions (1.2cm from left and right block edges)
    \node [sum] at ($(membrane.west) + (-1.2cm, 0)$) (sum) {};
    \node [circle, fill, inner sep=1.2pt] at ($(membrane.east) + (1.2cm, 0)$) (branch) {};

    % Outer Input / Output Terminals
    \node [coordinate, left=1.0cm of sum] (input_start) {};
    \node [coordinate, right=1.0cm of branch] (output) {};

    % Forward path
    \draw [->, line width=0.6pt] (input_start) -- node[pos=0.45, above] {$\delta i_{\mathrm{ext}}$} (sum);
    \draw [->, line width=0.6pt] (sum) -- (membrane);
    \draw [line width=0.6pt] (membrane) -- (branch);
    \draw [->, line width=0.6pt] (branch) -- node [pos=0.45, above] {$\delta v$} (output);

    % Feedback path
    \draw [->, line width=0.6pt] (branch) |- (ionic);
    \draw [->, line width=0.6pt] (ionic) -| (sum) 
      node [pos=0.25, above] {$\delta i_{\mathrm{int}}$};

    % Summation node signs
    \node at (sum.west) [anchor=north east, font=\scriptsize] {$+$}; % Below horizontal line
    \node at (sum.south) [anchor=north west, font=\scriptsize] {$-$}; % Right of vertical line

  \end{tikzpicture}
  \caption{Block diagram of neural dynamics.}
  \label{fig:neural_block_diagram}
\end{figure}

The key feature of the controller is its Right-Half-Plane (RHP) zero at 
$$s_z(v_0) = \frac{1-b + b v_0^2}{\tau(1-v_0^2)}$$
Since $0 < b < 1$, this zero is in the Right Half Plane for all $|v_0| < 1$.
We regard this RHP zero as the signature of mixed-feedback, namely, that the controller
is a difference of two gains: a stabilizing one, $\frac{1}{\tau s + b}$ (negative feedback), and a destabilizing one, $(1-v_0^2)$ (positive feedback).

It is the specific pole-zero configuration of the controller that forces the Hopf bifurcation 
of Fig.~\ref{fig:FN_closedloopHopf}.  As $|v_0|$ moves from $1$ toward $0$, the open-loop gain $1-v_0^2$ increases from $0$ 
to $1$, while the zero $s_z(v_0)$ moves from $+\infty$ to $\frac{1-b}{\tau}$. 
Thus, the closed-loop poles break away from the real interval $\left[-\frac{b}{\tau}, 0\right]$, branch out into the complex plane, and cross into the right half-plane (the Hopf bifurcation). 
The extent to which this locus is explored depends on the time constant $\tau$. 
A large $\tau$, 
% which is guaranteed by time-scale separation, plays a dual role: it 
increases the loop gain while shifting both the pole $s = -\frac{b}{\tau}$ and the zero $s_z(v_0)$  closer to the imaginary axis, facilitating right half-plane crossing, thus facilitating the spiking instability.

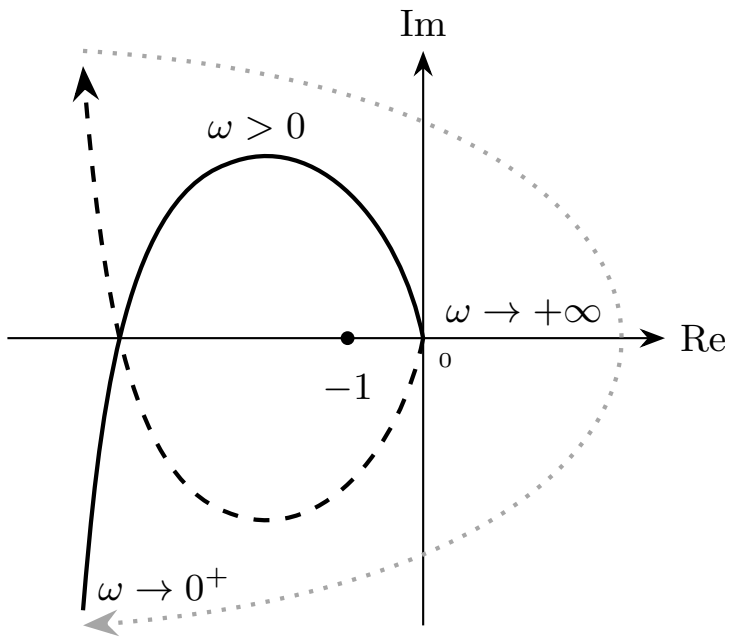
\begin{figure}[htbp]
\centering

\begin{tikzpicture}[scale=0.5, every node/.style={scale=0.8}, >=Stealth]

    % Axes (Plain white background, no grid)
    \draw[->, thin] (-5.5,0) -- (3.2,0) node[right, font=\small] {$\mathrm{Re}$};
    \draw[->, thin] (0,-3.8) -- (0,3.8) node[above, font=\small] {$\mathrm{Im}$};
    \node[font=\tiny, below right] at (0,0) {$0$};

    % Point -1 in black below the real axis
    \filldraw[black] (-1,0) circle (2.2pt);
    \node[below=4pt, font=\small\bfseries] at (-1,0) {$-1$};

    % -------------------------------------------------------------
    % Nyquist Locus for w > 0 (Solid Black Curve, 2 Arrows)
    % -------------------------------------------------------------
    \draw[black, thick]
        (-4.5, -3.6) 
        .. controls (-4.3, -1.2) and (-4.0, 1.8) .. (-2.6, 2.3)
        .. controls (-1.4, 2.8) and (-0.3, 1.5) .. (0,0);

    % -------------------------------------------------------------
    % Nyquist Locus for w < 0 (Dashed Black Curve, Restored Thickness)
    % -------------------------------------------------------------
    \draw[black, dashed, thick, ->]
        (0,0) 
        .. controls (-0.3, -1.5) and (-1.4, -2.8) .. (-2.6, -2.3)
        .. controls (-4.0, -1.8) and (-4.3, 1.2) .. (-4.5, 3.6);

    % -------------------------------------------------------------
    % Dashed Gray 'Closure to Infinity' Arc (Enclosing the locus)
    % -------------------------------------------------------------
    \draw[gray!70, dotted, thick, ->] 
        (-4.5, 3.8) .. controls (5, 3.2) and (5, -3.2) .. (-4.5, -3.8);
        % node[midway, right=4pt, font=\small, gray!90!black, align=left] 
        % {Closure arc at $\infty$\\[-2pt]($s = \epsilon e^{j\theta}, R \to \infty$)};

    % -------------------------------------------------------------
    % Frequency Annotations (Shifted positions)
    % -------------------------------------------------------------
    \node[black, font=\small, right] at (-4.5, -3.25) {$\omega \to 0^+$};
    \node[black, font=\small, right] at (0.1, 0.335) {$\omega \to +\infty$};
    \node[black, font=\small\bfseries] at (-2.2, 2.8) {$\omega > 0$};

\end{tikzpicture}

\caption{Structural Nyquist diagram of the return ration $L(v_0,s)$ for FitzHugh-Nagumo for voltages $v_0 < 1$ at which $L(v_0,s)$ is unstable.
The diagram is also valid for Hodgkin-Huxley (see Section \ref{sec:Nyquist_HH}), for $v_0$ in a selected region $\calV_u$ (Assumption \ref{assume:HH}), with the difference that the locus may converge to $0$ with a phase smaller than $-\frac{3}{2}\pi$.}
\label{fig:Nyquist_L}
\end{figure}

A simple yet insightful characterization can be derived from the Nyquist diagram of $L(v_0,s)$. 
For $|v_0| < 1$, as $\omega \to 0^+$, $L(v_0,j\omega)$ approaches infinity with a phase of $-\pi/2$. 
As $\omega$ increases, the phase decreases monotonically, converging to the origin as $\omega \to +\infty$ with a limiting phase of $-3\pi/2$. Therefore, for a sufficiently large time constant $\tau$, the complete Nyquist locus encircles $-1$ twice, as shown in Fig. \ref{fig:Nyquist_L}. Structurally, the loss of stability \emph{must occur} through a complex conjugate \emph{pair} of closed-loop 
poles crossing the imaginary axis, determining a Hopf bifurcation.

The Nyquist locus shows how the RHP zero of $K(v_0,s)$ provides the additional phase shift crucial for instability. Furthermore, the stability margin at the rest voltage 
$v_0$ is connected to the neuron's sensitivity to spiking. A reduced margin indicates resonant poles. This facilitates spiking because a small but fast (subthreshold) external current causes an overshoot in the voltage response, driving the voltage into the instability region (as in Fig. \ref{fig:FN_2x2_grid}).

\subsection{A linear control characterization of excitability}
% \label{sec:FN_nyquist}

The main proposal in this paper is to characterize excitability via two key features of the controller: the location of its RHP zero and the stability margin of the resting state, measured as the distance of the Nyquist plot of the return ratio $L(v_0,s)$ to the critical point $-1$. We propose that the RHP zero is a defining feature of any conductance-based model, whereas the type of bifurcation (Hopf in the example above) is model dependent. In the rest of the paper, we will show how the location of the RHP zero determines the type of excitability, whereas the stability margin determines the model sensitivity, that is, its propensity to spike.

\section{Conductance-based modelling}
\label{sec:CS}

\subsection{Linear feedback control representation}

\begin{figure}[htbp]
\centering
\begin{circuitikz}[scale=0.7, transform shape, font=\large] 
    % Main Rail and Input Voltage
    \draw (0,0) to[open, v<=$v$] (0,3); 
    \draw (0,3) -- (8.8,3);
    \draw (0,0) -- (8.8,0);
    
    % Membrane Capacitance Branch
    \draw (1,3) to[C, l=$C$] (1,0); 
    
    % Sodium Branch (Na)
    \draw (2.8,3) -- (2.8,2.5) 
                to[memristor] (2.8,1.3) 
                to[battery2, l=$E_\mathrm{Na}$] (2.8,0.05) -- (2.8,0); 
    \draw [->, >=stealth, thin, gray] (3.3, 2.3) -- (3.3, 1.5) node[midway, right, black] {$i_\mathrm{Na}$};
                
    % Potassium Branch (K)
    \draw (4.6,3) -- (4.6,2.5) 
                to[memristor] (4.6,1.3) 
                to[battery2, l=$E_\mathrm{K}$] (4.6,0.05) -- (4.6,0); 
    \draw [->, >=stealth, thin, gray] (5.1, 2.3) -- (5.1, 1.5) node[midway, right, black] {$i_\mathrm{K}$};
                
    % A-Current Branch (A)
    \draw (6.4,3) -- (6.4,2.5) 
                to[memristor] (6.4,1.3) 
                to[battery2, l=$E_A$] (6.4,0.05) -- (6.4,0); 
    \draw [->, >=stealth, thin, gray] (6.9, 2.3) -- (6.9, 1.5) node[midway, right, black] {$i_\mathrm{A}$};

    % Leakage Branch (L)
    \draw (8.2,3) -- (8.2,2.5) 
                to[memristor] (8.2,1.3) 
                to[battery2, l=$E_L$] (8.2,0.05) -- (8.2,0); 
    \draw [->, >=stealth, thin, gray] (8.7, 2.3) -- (8.7, 1.5) node[midway, right, black] {$i_L$};

    % External Applied Current
    \draw [->, >=stealth, thin, gray] (0,3.3) -- (1,3.3) node[midway, above, black] {$i_\mathrm{ext}$};
\end{circuitikz}
\caption{Connor-Stevens equivalent circuit.
The original Hodgkin-Huxley equivalent circuit is the same, with $i_\mathrm{A}$ removed.}
\label{fig:CS_circuit}
\end{figure}
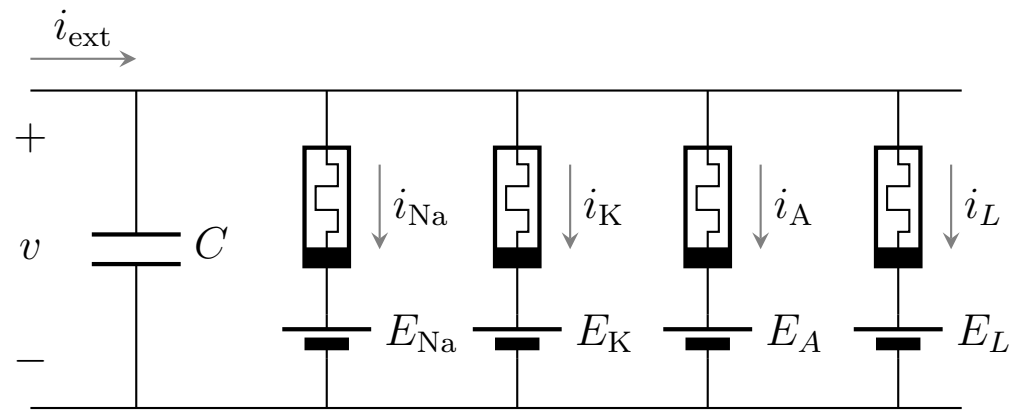

All conductance-based models have the circuit representation shown in Fig. \ref{fig:CS_circuit}: a linear capacitor, here regarded as the ``plant",
and a parallel bank of current sources, here regarded as the ``controller". Compared to the FitzHugh Nagumo model, only the model of the internal current differs: it is a sum of ohmic currents, each with a specific voltage-dependent conductance.

For the sake of illustration, the rest of the paper is concerned with the model of Connor and Stevens \cite{connor1977neural,abbot2001theoretical,connor1971voltage}, an early generalization of the seminal model of Hodgkin and Huxley \cite{hodgkin1952quantitative,Keener2009I,abbot2001theoretical}.
The equations read
\begin{subequations}
\begin{equation}
\begin{aligned}
C \dot{v} &= \underbrace{- i_\mathrm{L} - i_\mathrm{Na} - i_\mathrm{K}- i_{\mathrm{A}}}_{-i_\mathrm{int}} + i_{\mathrm{ext}}  \\
i_\mathrm{L} &= \bar{g}_{\mathrm{L}} (v \!-\! E_\mathrm{L}) \\
i_\mathrm{Na} &= \bar{g}_{\mathrm{Na}} m^3 h (v \!-\! E_{\mathrm{Na}}) \\
i_\mathrm{K} &= \bar{g}_{\mathrm{K}} n^4 (v \!-\! E_{\mathrm{K}}) \\
i_{\mathrm{A}} &= \bar{g}_{\text{A}} a^3 b (v - E_{\text{A}}),
\end{aligned}
\label{eq:HH_v}
\end{equation}
where $C>0$ is the membrance capacitance, the non-negative constants
$\bar{g}_{\mathrm{L}}$, $\bar{g}_{\mathrm{Na}}$, $\bar{g}_{\mathrm{K}}$, and $\bar{g}_{\mathrm{A}}$ are the so-called \emph{maximal conductances} (maximal amplification weights), and the constant $E_\mathrm{L}$, $E_\mathrm{Na}$, $E_\mathrm{K}$,  are the \emph{reverse potentials}.
These satisfy $ E_{\mathrm{K}} \le  v \le  E_{\mathrm{Na}}$. 
The ion-channel dynamics is represented by
\begin{equation}
\tau_x(v) \dot{x} = - x + x_\infty(v) 
\qquad x \in \mathcal{C} := \{m, n, h, a, b\},
\label{eq:HH_gating}
\end{equation}
\label{eq:HH}
which are first order lags whose time constants are voltage dependent and satisfy
\begin{equation}
\tau_m(v) \ll \tau_a(v) < \tau_b(v) < \tau_n(v) \simeq \tau_h(v), \quad \forall v.
\end{equation}
Each first order lag is driven by a voltage-dependent sigmoidal-like  nonlinearity $0 < x_\infty(v) < 1$, the so-called \emph{gating functions}.
The activation functions 
$m_\infty$, $n_\infty$, and $a_\infty$ are strictly increasing while 
the inactivation function $h_\infty$ and $b_\infty$
are strictly decreasing.

\end{subequations}

\begin{subequations}

The linearization of the controller at a given voltage $v_0$ has a very particular structure.
In the Laplace domain, each gating variable linearizes as a first-order lag
\begin{equation}
\delta x(s) = \frac{x'_\infty(v_0)}{\tau_x(v_0) s +1 } \delta v(s) \qquad x \in \mathcal{C}
\end{equation}
where $x'_\infty(v_0) = \left[\frac{\partial x_\infty(v)}{\partial v}\right]_{v=v_0}\!\!\!\!\!\!\!\!\!\!\!\!.$
The linearized internal current is therefore a linear combination of first-order lags
\begin{equation}
\begin{aligned}
\delta i_{\mathrm{int}}(s) 
& = \delta i_\mathrm{L}(s) + \delta i_\mathrm{Na}(s) 
+ \delta i_\mathrm{K}(s) + \delta i_\mathrm{A}(s) \\
& = K(v_0,s) \delta v(s) 
\end{aligned}
\end{equation}
for 
\begin{equation}
\begin{aligned}
K(v_0,s) 
&= \gamma_0(v_0) \ + \sum_{x\in\mathcal{C}} \frac{\gamma_x(v_0)}{\tau_x(v_0) s + 1} \\
\gamma_0(v_0) 
& = \bar{g}_{\mathrm{Na}} m_\infty^3(v_0) h_\infty(v_0) + \bar{g}_{\mathrm{K}} n_\infty^4(v_0) + \bar{g}_L \geq 0\\
\gamma_m(v_0) &= 3 \bar{g}_{\mathrm{Na}} m_\infty^2(v_0) h_\infty(v_0) (v_0 - E_{\mathrm{Na}}) \, m_\infty'(v_0) \leq 0 \\
\gamma_h(v_0) &= \bar{g}_{\mathrm{Na}} m_\infty^3(v_0) (v_0 - E_{\mathrm{Na}}) \, h_\infty'(v_0)  \geq 0\\
\gamma_n(v_0) &= 4 \bar{g}_{\mathrm{K}} n_\infty^3(v_0) (v_0 - E_{\mathrm{K}}) \, n_\infty'(v_0) \geq 0 \\
\gamma_a(v_0) &= 3 \bar{g}_{\text{A}} a_\infty^2(v_0) b_\infty(v_0) (v_0 - E_{\text{A}}) a_\infty'(v_0) \geq 0, \\
\gamma_b(v_0) &= \bar{g}_{\text{A}} a_\infty^3(v_0) (v_0 - E_{\text{A}}) b_\infty'(v_0) \leq 0.
\end{aligned}
\label{eq:G_ionic_HH}
\end{equation}
These inequalities assume that
$E_{\mathrm{K}} < v < E_{\mathrm{Na}}$. 
We also observe that the differential gains $\gamma_x$ in \eqref{eq:G_ionic_HH} are directly proportional to their corresponding maximal conductances and are nonzero whenever those conductances are nonzero.
The transfer function of the linearized Connor-Stevens model from $i_\mathrm{ext}$ to $v$ thus reads
\begin{equation}
\begin{aligned}
T(v_0,s) 
&= \frac{1}{C s + \gamma_0(v_0) +\sum\nolimits\limits_{x \in \mathcal{C}} \frac{\gamma_x(v_0)}{\tau_x(v_0) s + 1}}.
\end{aligned}
\label{eq:G_HH}
\end{equation}
\label{eq:HH_linearization}
\end{subequations}

\eqref{eq:G_HH} show remarkable similarities with \eqref{eq:FN_closedloop_TF}. 
If $\gamma_0(v_0)$ and  $\gamma_x(v_0)$ are all positive at all $v_0$, then $T(v_0,s)$ is a passive network. Each forced equilibrium $v_0$ is necessarily stable. In contrast, if $\bar{g}_{\mathrm{Na}}>0$, we have 
\begin{equation}
\gamma_m(v_0) < 0 \quad \mbox{ for all }
v_0.
\label{eq:HHpositive_feedback}
\end{equation}
In the linearization, the corresponding ionic channel provides \emph{positive feedback} $-\frac{|\gamma_m(v_0)|}{\tau_m(v_0)s +1}$.
This has a destabilizing effect for $\bar{g}_{\mathrm{Na}}$ sufficiently large.

\begin{figure}[htbp]
    \centering
    % Row 1
    \vspace{4mm}
    \begin{subfigure}[c]{0.46\columnwidth}
        \centering
        \includegraphics[width=\linewidth]{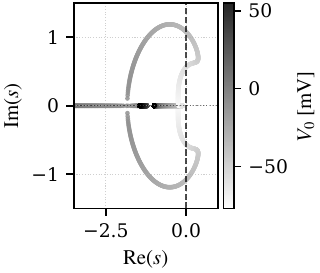}
        \caption{}
        \label{fig:HH_closedloopHopf}
    \end{subfigure}%
    \begin{subfigure}[c]{0.46\columnwidth}
        \centering
        \includegraphics[width=\linewidth]{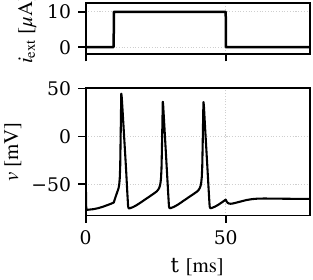}
        \caption{}
        \label{fig:HH_voltagetrajectories}
    \end{subfigure}

    \vspace{2ex}

    % Row 2
    \begin{subfigure}[c]{0.46\columnwidth}
        \centering
        \includegraphics[width=\linewidth]{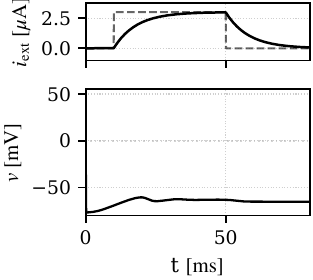}
        \caption{}
        \label{fig:HH_filtered_slow}
    \end{subfigure}%
    \begin{subfigure}[c]{0.46\columnwidth}
        \centering
        \includegraphics[width=\linewidth]{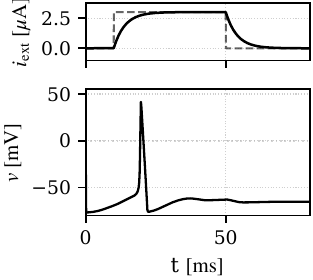}
        \caption{}
        \label{fig:HH_filtered_fast}
    \end{subfigure}

    \caption{Hodgkin-Huxley model: (a) Poles of $T_{\mathrm{HH}}(v_0,s)$ at $-80 \leq v_0 \leq 55$, (b) Spiking behavior for $i_{\mathrm{ext}}=10\,\mu\mathrm{A}$, and (c)–(d) Spiking behavior for filtered $i_\mathrm{ext}(t) \leq 3\,\mu\mathrm{A}$ with time constants $\tau_f = 4$ and $\tau_f = 2$.}
    \label{fig:HH_2x2_grid}
\end{figure}

The complexity of the closed-loop transfer function  does not allow for a parametric analysis of the denominator roots, as in the FitzHugh-Nagumo case.  
However, these can be analyzed numerically, as shown in Fig. \ref{fig:HH_closedloopHopf}, which adopts
the 
classical setting of Hodgkin-Huxley neuron \cite{hodgkin1952quantitative}, adapted to modern membrane potential coordinates. Specifically, we take $\bar{g}_{\text{A}} = 0$ while the other parameters are from  \cite[Table 3]{hodgkin1952quantitative}.
As in the FitzHugh-Nagumo case,
Figs. \ref{fig:HH_filtered_fast} and \ref{fig:HH_filtered_slow}
shows the different response of the neuron to fast and slow varying external currents, due to overshoot.

\subsection{Mixed-feedback and excitability}

A remarkable property of conductance-based models is that, regardless of their dimension, the linearized controller $K(v_0,s)$ can {\it always}
be written as the difference of two relaxation systems. Relaxation systems \cite{Willems1972Part2} are {\it positive} combinations of first-order lags. 
A difference of relaxation systems writes as
\begin{equation}
K(v_0,s) = - P(v_0,s) + N(v_0,s),
\label{eq:K_mixed}
\end{equation}
where both $P$ and $N$ are relaxation systems. We call $N(v_0,s)$ a negative feedback controller, because it is passive and can never destabilize a passive membrane. Note that this controller has the dimension of a differential conductance as a ratio $\frac{\delta i}{\delta v}$.  Negative feedback is thus also synonym of positive differential conductance. For the same reason, we call $-P(v_0,s)$ a positive feedback controller.
In Connor-Stevens model model, $N(v_0,s)$ collects the positive differential conductances, that is,  the sodium inactivation $h$, potassium activation $n$, type-A potassium activation $a$ and passive leak. In contrast, $-P(v_0,s)$ aggregates the negative differential conductances, that is, the sodium activation $m$ and the type-A potassium inactivation $b$. In  FitzHugh-Nagumo model, $N(v_0,s)$ simplifies to a single first-order lag whereas $P(v_0,s)$ simplifies to a static gain. 

This mixed feedback architecture is fundamental to neural excitability \cite{sepulchre2019,sepulchre2022,ribar2021}.
The fast positive feedback of sodium activation \emph{enables} the instability required for spike generation, while the remaining networks at slower time scale dictates \emph{how} stability is lost and restored. This shapes both the voltage range $\mathcal{V}_u$ of unstable linearizations and the bifurcation mechanism governing spike initiation.

The proposal in the present paper is to observe that the mixed-feedback architecture forces a
structural \emph{RHP zero} in the (linearized) open-loop admittance $K(v_0,s)$, and that the location of this zero shapes the Nyquist plot of the return ratio, thereby determining the type of bifurcation and sensitivity at equilibrium.  

\begin{proposition}
\label{rem:nonmin_phase_zero}
\emph{RHP zero in mixed relaxation networks.} 
Let $0 < \alpha \in \real$ and $\beta_i \in \real$. Consider $\tau_f>0$ and $\tau_i > 0$ for all $1 \leq i \leq n$. The transfer function
\begin{equation}
- \frac{\alpha}{\tau_f s+1} + \sum_{i=1}^n \frac{\beta_i}{\tau_i s + 1}
\end{equation}
always has a RHP zero if $\sum_{i=1}^n \beta_i > \alpha$ and $\tau_f$ is sufficiently small. 
\end{proposition}

\begin{proof} 
Compute the polynomial at the numerator. 
Its static coefficient 
\begin{equation}
c_0 = \sum_{i=1}^n \beta_i - \alpha > 0
\end{equation} by assumption. 
The coefficient that multiplies the largest monomial $s^n$ reads
\begin{equation}
\begin{aligned}
c_n &= -\alpha \prod_{j=1}^n \tau_j + \tau_f \sum_{i=1}^n \left( \beta_i \prod_{j \neq i}^n \tau_j \right) \\ &= \left( -\alpha + \tau_f \sum_{i=1}^n \frac{\beta_i}{\tau_i} \right) \prod_{j=1}^n \tau_j < 0
\end{aligned}
\end{equation}
for $\tau_f$ sufficiently small. Under these conditions, the numerator is equal to $c_0>0$ for $s=0$ and
goes to $-\infty$ for $s\to\infty$. Thus, there is at least one real zero at $s \in (0, \infty)$. This zero cannot be canceled by a pole, as all open-loop poles lie strictly in the left half-plane.
That is, a \emph{positive zero is guaranteed to exist} if $\alpha$ satisfies
\begin{equation}
\tau_f \sum_{i=1}^n \frac{\beta_i}{\tau_i} < \alpha < \sum_{i=1}^n \beta_i.
\end{equation}
\end{proof}

\section{Mixed-feedback regulation of excitability}
\label{sec:threshold_regulation}

Early in the development of neuronal modeling, Hodgkin classified different {\it types} of excitability \cite{prescott2014excitability}. The traditional approach in neurodynamics
has been to ground the type of excitability in bifurcation analysis: the Hopf bifurcation in Hodgkin-Huxley
model is the landmark of Type II excitability, which imposes a minimal spiking frequency near threshold, whereas the saddle-node
bifurcation of Connor-Stevens model is the landmark of Type I excitability, which allows for arbitrarily low spiking frequency.

In this section, we revisit this classification in terms of the Nyquist analysis of the mixed-feedback controller, that is,
an input-output property of the total differential conductance of the internal current, that can be continuously adjusted in any
conductance-based model by neuromodulation.

Our only analysis tool is the  Nyquist plot of  the return ratio \eqref{eq:return_ratio}
\begin{equation}
L(v_0,s) 
= 
\frac{1}{Cs}\underbrace{\left(\gamma_0(v_0) \ + \sum_{x\in\mathcal{C}} \frac{\gamma_x(v_0)}{\tau_x(v_0) s + 1}\right)}_{K(v_0,s)}.
\end{equation}
which is a curve that can be drawn by standard numerical tools for an arbitrary conductance-based model. We apply this
analysis to Connor-Stevens model, which is known to only exhibit Type II excitability when $\bar{g}_A=0$ (Hodgkin-Huxley model),
and that can transition to type I excitability under the regulatory action of the type-A potassium current.

\subsection{Type II excitability}
\label{sec:Nyquist_HH}

For $\bar{g}_\mathrm{A} = 0$,  the only source of positive feedback is the fast activation of the sodium current.
The shape of the Nyquist diagram 
depends on the relative strength of the four
gains $\gamma_0$, $\gamma_m$, $\gamma_h$, and $\gamma_n$
and on the relative time-scale of each lag.
A structural result can be derived under the following basic assumption.
\begin{assume}
\label{assume:HH}
There exists a voltage region $\calV_u$ and a constant $0 \leq \alpha < 1$ such that, for all $v_0 \in \calV_u$, 
\begin{subequations}
\begin{align}
\gamma_0(v_0) + \gamma_m(v_0) + \gamma_h(v_0) + \gamma_n(v_0) & > 0
\label{eq:HH_DC} \\
\gamma_0(v_0) + \gamma_m(v_0) + \alpha \left( \gamma_h(v_0) + \gamma_n(v_0) \right) & < 0 
\label{eq:HH_instability}
\end{align} 
\label{eq:HH_gains}
\end{subequations} 
\end{assume}
\eqref{eq:HH_DC} guarantees a positive DC gain $K(v_0,0) > 0$.
\eqref{eq:HH_instability} ensures that
the positive feedback gain $|\gamma_m(v_0)|$ 
is \emph{sufficiently strong} to induce instability, overcoming static resistive dissipation $\gamma_0(v_0)$ and $\alpha$-attenuated dissipation of the slow relaxation terms $n$ and $h$.

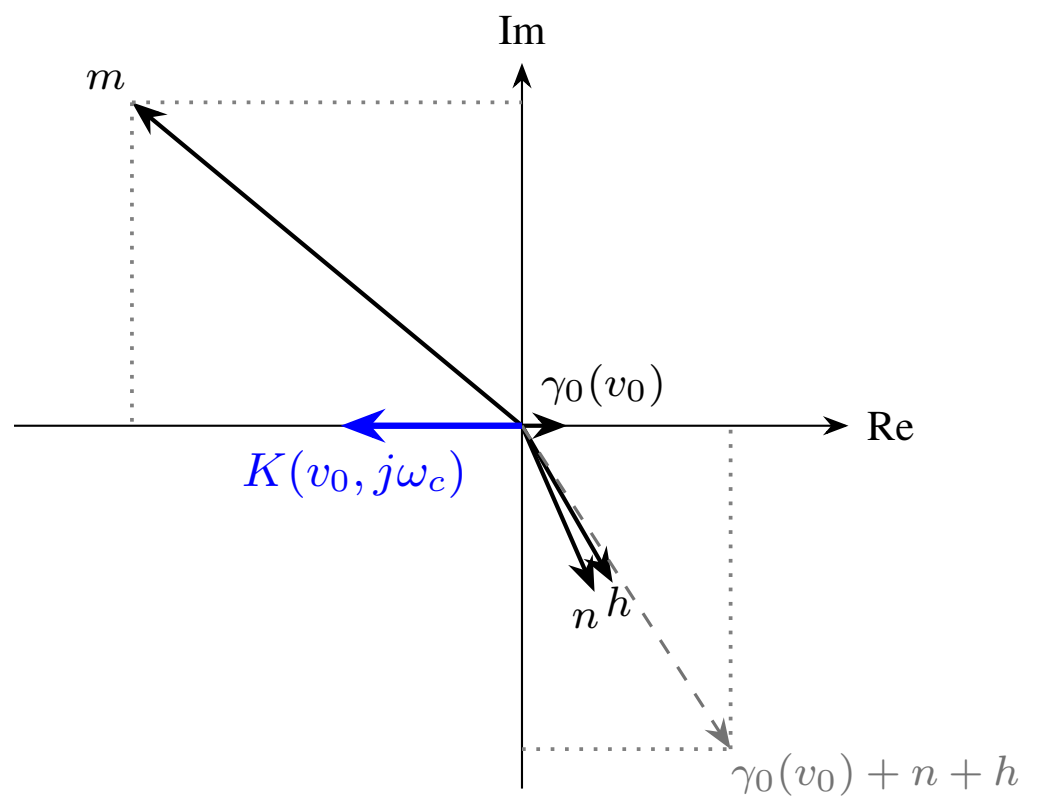
\begin{figure}[htbp]
\centering
\begin{tikzpicture}[scale=1.2, >=Stealth, font=\footnotesize]

    % Axes (UltratHIN lines)
    \draw[->, thin] (-2.8,0) -- (1.8,0);
    \draw[->, thin] (0,-2.0) -- (0,2.0);

    % Vector Endpoint Coordinates
    \coordinate (O)  at (0, 0);
    \coordinate (G0) at (0.25, 0.0);       % \gamma_0 (length 0.25)
    \coordinate (H)  at (0.50, -0.866);    % h (length 1.0)
    \coordinate (N)  at (0.40, -0.9165);   % n (length 1.0)
    \coordinate (Vs) at (1.15, -1.7825);   % Sum (\gamma_0 + h + n)
    \coordinate (Vm) at (-2.15, 1.7825);   % m
    \coordinate (K)  at (-1.00, 0.0);      % K_ion (-1, 0)
    \coordinate (V0plusH) at (0.75, -0.866); % Tip-to-tail intermediate point

    % -------------------------------------------------------------
    % Vector Paths
    % -------------------------------------------------------------
    \draw[->, black, line width=0.8pt] (O) -- (G0);
    \draw[->, black, line width=0.8pt] (O) -- (H);
    \draw[->, black, line width=0.8pt] (O) -- (N);
    \draw[->, black, line width=0.8pt] (O) -- (Vm);
    \draw[->, gray!90!black, dashed, line width=0.6pt] (O) -- (Vs);
    \draw[->, blue, line width=1.2pt] (O) -- (K);

    % Construction & Alignment Dotted Lines
    % \draw[black, dotted, thin] (G0) -- (V0plusH) -- (Vs);
    \draw[gray, dotted, line width=0.7pt] (-2.15, 1.7825) -- (0, 1.7825);
    \draw[gray, dotted, line width=0.7pt] (0, -1.7825) -- (1.15, -1.7825);
    \draw[gray, dotted, line width=0.7pt] (1.15, -1.7825) -- (1.15, 0);
    \draw[gray, dotted, line width=0.7pt] (-2.15, 1.7825) -- (-2.15, 0);

    % -------------------------------------------------------------
    % Labels with Explicit Coordinates (for easy micro-adjustments)
    % -------------------------------------------------------------
    \node[right] at (1.80, 0.00) {Re};
    \node[above] at (0.00, 2.00) {Im};

    \node[above=1pt] at (0.45, -0.025) {$\gamma_0(v_0)$};
    \node[right=2pt] at (0.3, -0.966) {$h$};
    \node[below=2pt] at (0.350, -0.85) {$n$};
    \node[above left] at (-2.08, 1.75) {$m$};
    
    \node[below right, gray!90!black, align=left] at (1.050, -1.7) {$\gamma_0(v_0) + n + h$};
    \node[above left, blue, font=\small] at (-0.200, -0.50) {$K(v_0, j\omega_c)$};

\end{tikzpicture}
\caption{Hodgkin-Huxley $K(v_0,j \omega_c)$ as sum
of four complex numbers. The letters $n$, $h$, $m$ indicate the numbers $\frac{\gamma_x(v_0)}{\tau_x(v_0) j\omega_c  + 1}$, for $x\in\{m,h,n\}$.}
\label{fig:HH_structural_nyquist}
\end{figure}

For all $v_0$ and $\omega \in [0,\infty)$ the individual
loci of $\gamma_0(v_0)$, $\frac{\gamma_h(v_0)}{\tau_h(v_0) j\omega  + 1}$, and $\frac{\gamma_n(v_0)}{\tau_n(v_0) j\omega  + 1}$ lie in the right half-plane, as illustrated in
Fig. \ref{fig:HH_structural_nyquist}.
The loci of these first order lags start on the positive real axis for $\omega \to 0^+$. As $\omega$ increases, the phase decreases monotonically, ending at  $0$ with phase $-\frac{\pi}{2}$ for $\omega \to \infty$. Therefore, the locus of their sum follows the same qualitative behavior. 
In contrast, because $\gamma_m(v_0) \leq 0$, the locus of $\frac{\gamma_m(v_0)}{\tau_m(v_0) j\omega  + 1}$ lies entirely within the left half-plane. It starts on the negative real axis as $\omega \to 0^+$ and, as $\omega$ increases, its phase decreases monotonically towards $-\frac{3\pi}{2}$. The locus converges to $0$ with phase $-\frac{3\pi}{2}$ for $\omega \to \infty$. 

\begin{figure}[t]
    \centering
    \includegraphics[width=0.49\columnwidth]{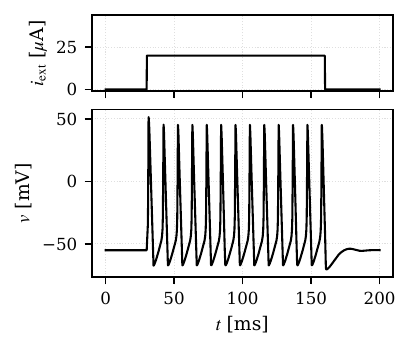}
    \includegraphics[width=0.49\columnwidth]{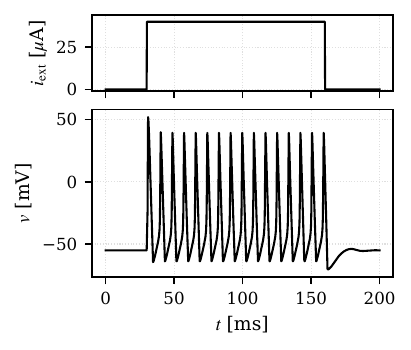}
    \includegraphics[width=0.49\columnwidth]{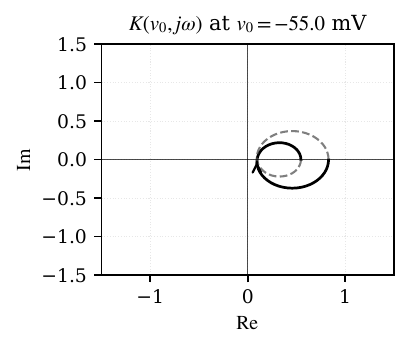}
    \includegraphics[width=0.49\columnwidth]{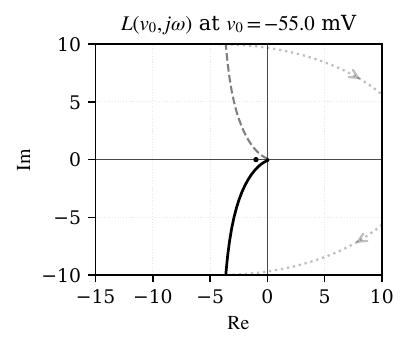}
    \includegraphics[width=0.49\columnwidth]{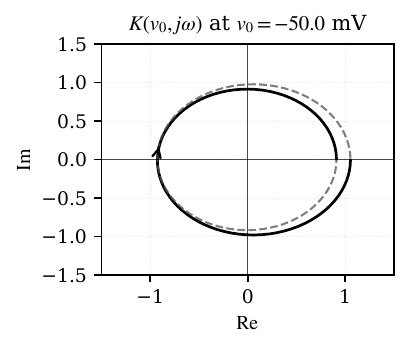}
    \includegraphics[width=0.49\columnwidth]{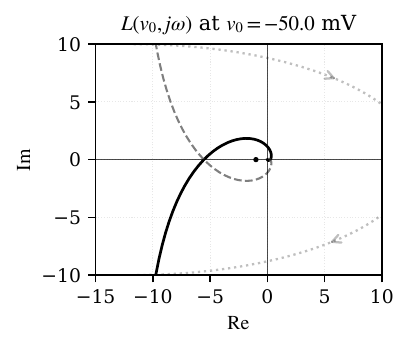}
    \vspace*{-5mm}
      \caption{Connor--Stevens model for $\bar{g}_\mathrm{A} = 0$ (Hodgkin-Huxley). 
  \textbf{Top:} voltage response to sustained currents $i_{\mathrm{ext}} = 20\,\mu\mathrm{A}$ (left) and $i_{\mathrm{ext}} = 40\,\mu\mathrm{A}$ (right).
  \textbf{Middle and bottom:} Nyquist loci of $K(v_0, j\omega)$ and $L(v_0, j\omega)$, at sub-threshold voltage $v_0 = -55\,\mathrm{mV}$ (middle) and super-threshold voltage $v_0 = -50\,\mathrm{mV}$ (bottom). 
  This illustrates the transition to instability typical of a Hopf bifurcation
  (Type-II excitability).}
  \label{fig:HH_nyquist}
\end{figure}

These properties combined with Assumption \ref{assume:HH} provide the ground for
a Hopf bifurcation. For any $v_0 \in \calV_u$:
\begin{itemize}
\item By \eqref{eq:HH_DC}, the locus of
$K(v_0,j\omega)$ originates on the real axis for $\omega \to 0^+$. 
\item Under sufficient 
time-scale separation $\tau_{m}(v_0) \ll \tau_h(v_0)$ and $\tau_{m}(v_0) \ll \tau_n(v_0)$, \eqref{eq:HH_instability} guarantees that 
$K(v_0,j\omega_c)$ crosses the negative real axis 
at a critical frequency $\omega_c$. At this frequency the positive
imaginary component of 
$\frac{\gamma_m(v_0)}{\tau_m(v_0) j\omega_c  + 1}$ 
exactly cancels the
negative imaginary part of the sum $\gamma_0(v_0) + \frac{\gamma_h(v_0)}{\tau_h(v_0) j\omega_c  + 1} + \frac{\gamma_n(v_0)}{\tau_n(v_0) j\omega_c  + 1}$, taking advantage of the magnitude attenuation of the slow channels and \eqref{eq:HH_instability}. This is illustrated in Fig. \ref{fig:HH_structural_nyquist}.
\item Due to time scale separation, $\frac{\gamma_m(v_0)}{\tau_m(v_0) j\omega_c  + 1}$ dominates the high-frequency response, ensuring that the phase of $K(v_0,j\omega)$ decreases without re-crossing of the negative real axis. 
\item Finally, $K(v_0,j\omega)$ converges to $\gamma_0(v_0)$ as $\omega \to \infty$. 
\end{itemize}
A summarizing locus is provided in Fig. \ref{fig:HH_nyquist}. Spiking simulations and loci are derived for gating functions, voltage-dependent time constants, and parameters from \cite{connor1977neural}. Only the leak reverse potential has been changed to $E_\mathrm{L} = -47$mV.

 The Nyquist locus of $L(v_0,s)$ is obtained by introducing a $-\frac{\pi}{2}$ phase lag and a magnitude scaling of $\frac{1}{C\omega}$. Thus, $L(v_0, j\omega)$ approaches infinity with phase $-\frac{\pi}{2}$ for $\omega \to 0^+$, crosses the negative real axis with a monotonically decreasing phase, and converges to $0$ as $\omega \to \infty$. 
As in the FitzHugh-Nagumo case, for a sufficiently small capacitance $C$, the loss of stability \emph{must occur} through a complex conjugate \emph{pair} of closed-loop poles crossing the imaginary axis, as shown in Fig. \ref{fig:HH_nyquist}. This Hopf bifurcation underlies Type II excitability \cite{prescott2014excitability}. The non-zero imaginary part of the right-half plane poles is directly related the the minimum frequency of spiking, typical of Type II excitability.

\subsection{Type I excitability}
\begin{figure}[t]
    \centering
    \includegraphics[width=0.49\columnwidth]{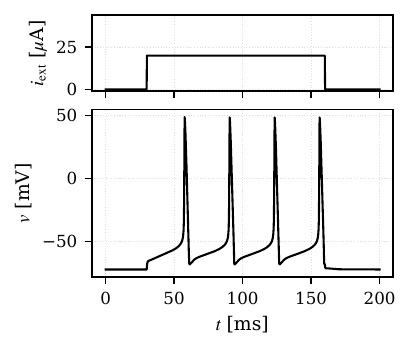}
    \includegraphics[width=0.49\columnwidth]{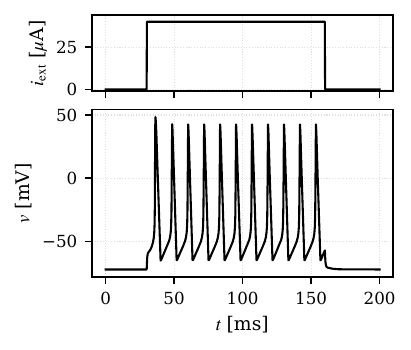}
    \includegraphics[width=0.49\columnwidth]{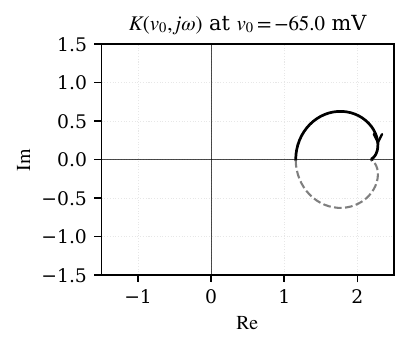}
    \includegraphics[width=0.49\columnwidth]{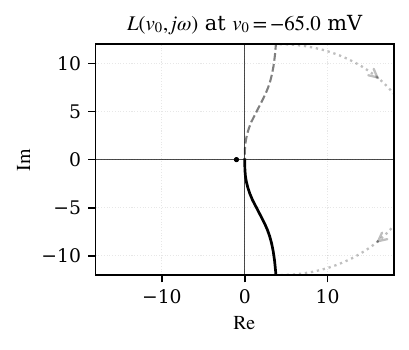}
    \includegraphics[width=0.49\columnwidth]{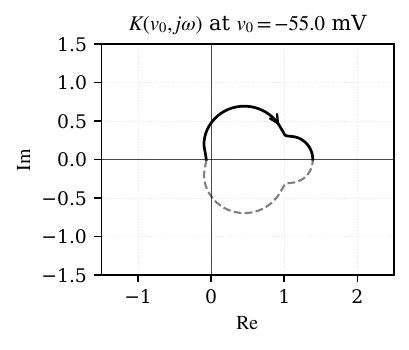}
    \includegraphics[width=0.49\columnwidth]{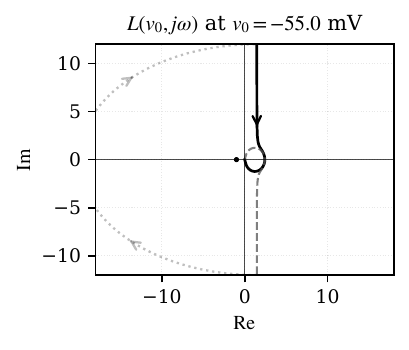}
    \vspace*{-5mm}
      \caption{Connor--Stevens model for $\bar{g}_\mathrm{A} = 47.7$. 
  \textbf{Top:} voltage response to sustained currents $i_{\mathrm{ext}} = 20\,\mu\mathrm{A}$ (left) and $i_{\mathrm{ext}} = 40\,\mu\mathrm{A}$ (right),
  demonstrating current-dependent firing frequency modulation characteristic of Type-I excitability.
  \textbf{Middle and bottom:} Nyquist loci of $K(v_0, j\omega)$ and $L(v_0, j\omega)$, at sub-threshold voltage $v_0 = -65\,\mathrm{mV}$ (middle) and super-threshold voltage $v_0 = -55\,\mathrm{mV}$ (bottom). 
  This illustrates the transition to instability of a saddle-node bifurcation
  (Type-I excitability).}
  \label{fig:CS}
\end{figure}

For $\bar{g}_A > 0$, the transient A-current introduces a new source of positive feedback in the slow time scale $\tau_b$ of type-A potassium inactivation.
The fast $a$-activation supplies fast \emph{negative feedback} that can balance sodium activation. The slower $b$-inactivation yields a negative differential conductance ($\gamma_b < 0$), providing slow \emph{positive feedback} which counteracts the stabilizing influence of $h$ and $n$, altering the low-frequency features of the ionic admittance $K(v_0, s)$. 
This is illustrated numerically in Fig. \ref{fig:CS},
derived for gating functions, time constants, and parameters from \cite{connor1977neural} (besides $E_\mathrm{L} = -47$mV).

In contrast to the case $\bar{g}_\mathrm{A}= 0$, the slow positive feedback alters the Nyquist locus, bringing the number of clockwise encirclements of $-1$ to $1$. This is the signature of saddle-node bifurcation. This bifurcation  is at the core of Type-I excitability \cite{prescott2014excitability}, allowing the neuron to sustain arbitrarily low firing frequencies as a function of stimulus strength. 

The Nyquist analysis demonstrates
how the slow conductances shape the Nyquist locus and affects how the closed-loop poles cross the imaginary axis.  
In contrast to bifurcation analysis, the modulation of the Nyquist plot is a direct measure of the input-output differential conductance
of the total ionic current. It is a robust analysis tool in the same way as in classical linear control analysis. Yet it is consistent with the results of neurodynamics and does not prevent a discrete classification: there is no bifurcation without positive feedback current; with $g_A=0$, only Type II excitability is possible; etc...

\section{Mixed-feedback regulation of sensitivity}
\label{sec:sensitivity_regulation}

As in classical robust linear theory, Nyquist analysis provides a framework for evaluating the sensitivity of spiking. Parametric and dynamic uncertainties correspond to graphical perturbations of the Nyquist locus. Therefore, closed-loop instability, and hence spiking, persists under uncertainties that preserve the encirclement of the critical point $-1$. Even more importantly, the distance of the Nyquist plot to the critical point is a direct quantitative measure of sensitivity, that is, the propensity to spike.

\begin{figure}[htbp]
    \centering
    \includegraphics[width=0.49\columnwidth]{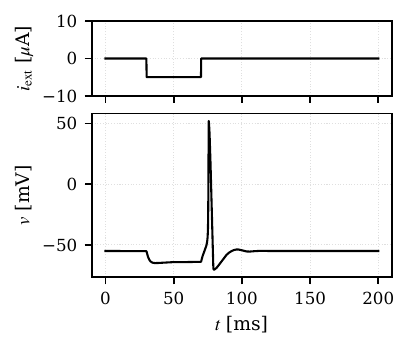}
    \includegraphics[width=0.49\columnwidth]{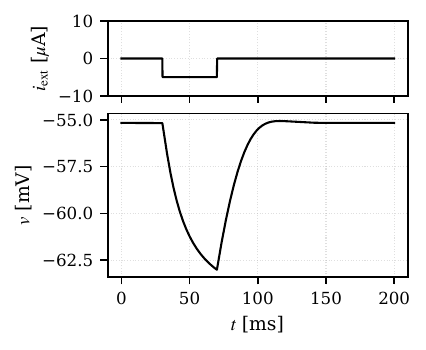}
    \includegraphics[width=0.49\columnwidth]{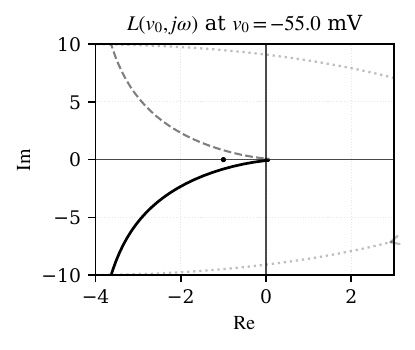}
    \includegraphics[width=0.49\columnwidth]{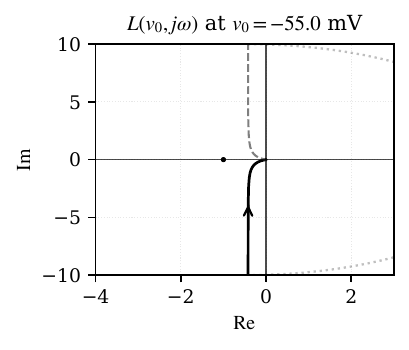}
    \vspace*{-5mm}
      \caption{Connor--Stevens model for $\bar{g}_\mathrm{A} = 0$ (Hodgkin--Huxley). 
Left: $C=1$; Right: $C=10$. The small stability margin for $C=1$ causes a step-response overshoot when hyperpolarization is released, which triggers a spike.}
  \label{fig:HH_sensitivity}
\end{figure}

Figs.~\ref{fig:HH_filtered_slow} and \ref{fig:HH_filtered_fast} demonstrate that small, fast-varying external inputs produce a voltage overshoot sufficient to trigger a spike. This is not an artifact but a consequence of the Nyquist locus. Fig. \ref{fig:HH_sensitivity} provides a clear illustration. Both Nyquist plots refer to the Connor-Stevens model for $\bar{g}_{\mathrm{A}}=0$ (Hopf bifurcation, resonant Type II excitability), linearized at the resting potential $v_0 \simeq -55\mathrm{mV}$ for $i_\mathrm{ext} = 0$. Both plots are based on parameters from \cite{connor1977neural} (besides $E_\mathrm{L} = -47$mV). Their difference lies in the value of the membrane capacitance: the left plot corresponds to $C=1$, while the right plot uses $C=10$. From \eqref{eq:return_ratio}, a larger capacitance scales the Nyquist plot, therefore the stability margin of the closed-loop system. 

For $C=1$, the stability margin is reduced. Fast-changing external currents cause an overshoot in the voltage response, triggering spiking. This is well illustrated by the \emph{post-inhibitory spike} in the left column of Fig. \ref{fig:HH_sensitivity}. The fast release of an external hyperpolarizing current is sufficient to trigger a spike. This phenomenon is called post-inhibitory facilitation in the literature \cite{dodla2022}.
The $C=10$ case is characterized by a large stability margin. For these parameters, the same external currents generate a well-damped voltage response with no overshoot. As a result, the post-inhibitory spike is lost even if the neuron preserves its Type II excitability
(and both would spike for a sufficiently large external current,
as in Fig. \ref{fig:HH_nyquist}).

\section{Conclusion}

The control design studied in this paper is standard: the regulation of a first-order passive plant by a non-minimum
phase controller that includes a RHP zero due as a consequence of being the difference of a fast relaxation system (fast positive feedback) and a slow relaxation system (slow negative feedback). We have proposed
that this linear control architecture is the core mechanism that regulates spiking. The regulation of spiking can be understood from classical root locus and Nyquist analysis of the return ratio. This analysis applies to arbitrary conductance-based models, regardless of their dimension.
The analysis suggests the potential of control theory to address inherent limitations of bifurcation analysis: the need for a methodology
that is does not require reduced-order models, that models uncertainty and modulation as a continuous deformation of an input-output differential conductance, and that
``tunes" rather than ``classifies". Future work will include the application of the methodology to challenging questions of
neuronal excitability, such as the tuning of rebound excitability by additional slow currents and the extension of the 
methodology to burst (rather than spike) excitability.

\addtolength{\textheight}{-12cm}   % This command serves to balance the column lengths

%%%%%%%%%%%%%%%%%%%%%%%%%%%%%%%%%%%%%%%%%%%%%%%%%%%%%%%%%%%%%%%%%%%%%%%%%%%%%%%%

%%%%%%%%%%%%%%%%%%%%%%%%%%%%%%%%%%%%%%%%%%%%%%%%%%%%%%%%%%%%%%%%%%%%%%%%%%%%%%%%


% Later

\section*{Acknowledgment}
Gemini 3.6 was used to generate code for diagrams and plots, to assist with standard mathematical derivations (such as the computation of linearized dynamics), and for language editing of selected paragraphs. The authors reviewed and edited the content as needed and take full responsibility for the content of the publication.

%%%%%%%%%%%%%%%%%%%%%%%%%%%%%%%%%%%%%%%%%%%%%%%%%%%%%%%%%%%%%%%%%%%%%%%%%%%%%%%%

\balance
\bibliographystyle{IEEEtran}
\bibliography{FF+RS_bibliography,FF+RS_bibliography2,FF+RS_bibliography3}

@ARTICLE{sepulchre2022,
  author={Sepulchre, Rodolphe},
  journal={Proceedings of the IEEE}, 
  title={Spiking Control Systems}, 
  year={2022},
  volume={110},
  number={5},
  pages={577-589},
  doi={10.1109/JPROC.2022.3163926}}

@article{sepulchre2019,
	author = {Sepulchre, R. and Drion, G. and Franci, A.},
	doi = {10.1146/annurev-control-053018-023708},
	issn = {2573-5144},
	journal = {Annual Review of Control, Robotics, and Autonomous Systems},
	month = may,
	number = {1},
	pages = {89--113},
	title = {Control {{Across Scales}} by {{Positive}} and {{Negative Feedback}}},
	volume = {2},
	year = {2019}}

@article{ribar2021,
	author = {Ribar, Luka and Sepulchre, Rodolphe},
	doi = {10.1109/MCS.2021.3107560},
	journal = {IEEE Control Systems Magazine},
	number = {6},
	pages = {34-63},
	title = {Neuromorphic Control: Designing Multiscale Mixed-Feedback Systems},
	volume = {41},
	year = {2021}}

@article{fitzhugh1961impulses,
  author    = {FitzHugh, R.},
  title     = {Impulses and Physiological States in Theoretical Models of Nerve Membrane},
  journal   = {Biophysical Journal},
  volume    = {1},
  number    = {6},
  pages     = {445--466},
  year      = {1961},
  publisher = {Elsevier},
  doi       = {10.1016/s0006-3495(61)86902-6}
}

@Inbook{sherwood2014fitzhughnagumo,
  author    = {Sherwood, W. E.},
  title     = {FitzHugh--Nagumo Model},
  editor    = {Jaeger, Dieter and Jung, Ranu},
  booktitle = {Encyclopedia of Computational Neuroscience},
  publisher = {Springer},
  address   = {New York, NY},
  year      = {2014},
  doi       = {10.1007/978-1-4614-7320-6_147-1},
}

@incollection{dodla2022,
  author    = {Dodla, Ramana and Svirskis, Gytis and Rinzel, John},
  title     = {Postinhibitory Rebound and Facilitation},
  booktitle = {Encyclopedia of Computational Neuroscience},
  editor    = {Jaeger, Dieter and Jung, Ranu},
  year      = {2022},
  publisher = {Springer Berlin Heidelberg},
  address   = {Berlin, Heidelberg},
  pages     = {2847--2854},
  isbn      = {978-1-0716-1006-0},
  doi       = {10.1007/978-1-0716-1006-0_152}
}

@article{franci2018,
  author    = {Franci, Alessio and Drion, Guillaume and Sepulchre, Rodolphe},
  title     = {Robust and tunable bursting requires slow positive feedback},
  journal   = {Journal of Neurophysiology},
  volume    = {119},
  number    = {3},
  pages     = {1222--1234},
  year      = {2018},
  doi       = {10.1152/jn.00804.2017},
  pmid      = {29357476},
  url       = {https://journals.physiology.org/doi/full/10.1152/jn.00804.2017}
}

@article{drion2012,
  author    = {Drion, Guillaume and Franci, Alessio and Seutin, Vincent and Sepulchre, Rodolphe},
  title     = {A Novel Phase Portrait for Neuronal Excitability},
  journal   = {PLOS ONE},
  volume    = {7},
  number    = {8},
  pages     = {e41806},
  year      = {2012},
  month     = {08},
  publisher = {Public Library of Science},
  doi       = {10.1371/journal.pone.0041806},
  url       = {https://doi.org/10.1371/journal.pone.0041806}
}

@inproceedings{sepulchre2019Control,
  author    = {Sepulchre, Rodolphe and O'Leary, Timothy and Drion, Guillaume and Franci, Alessio},
  title     = {Control by neuromodulation: A tutorial},
  booktitle = {2019 18th European Control Conference (ECC)},
  year      = {2019},
  pages     = {483--497},
  address   = {Naples, Italy},
  doi       = {10.23919/ECC.2019.8795686}
}

@article{marder2014,
  author    = {Marder, Eve and O'Leary, Tim and Shruti, Sonal},
  title     = {Neuromodulation of Circuits with Variable Parameters: Single Neurons and Small Circuits Reveal Principles of State-Dependent and Robust Neuromodulation},
  journal   = {Annual Review of Neuroscience},
  volume    = {37},
  number    = {1},
  pages     = {329--346},
  year      = {2014},
  doi       = {10.1146/annurev-neuro-071013-013958},
  URL       = {https://annualreviews.org}
}

@article{franci2012,
  author    = {Franci, Alessio and Drion, Guillaume and Sepulchre, Rodolphe},
  title     = {An Organizing Center in a Planar Model of Neuronal Excitability},
  journal   = {SIAM Journal on Applied Dynamical Systems},
  volume    = {11},
  number    = {4},
  pages     = {1698--1722},
  year      = {2012},
  publisher = {SIAM},
  doi       = {10.1137/110859590}
}

@ARTICLE{Nagumo1962pulsetransmission,
  author={Nagumo, J. and Arimoto, S. and Yoshizawa, S.},
  journal={Proceedings of the IRE}, 
  title={An Active Pulse Transmission Line Simulating Nerve Axon}, 
  year={1962},
  volume={50},
  number={10},
  pages={2061-2070},
  doi={10.1109/JRPROC.1962.288235}}

@article{mauro1970subthreshold,
  author    = {Mauro, A. and Conti, F. and Dodge, F. and Schor, R.},
  title     = {Subthreshold Behavior and Phenomenological Impedance of the Squid Giant Axon},
  journal   = {The Journal of General Physiology},
  volume    = {55},
  number    = {4},
  pages     = {497--523},
  year      = {1970},
  publisher = {Rockefeller University Press},
  doi       = {10.1085/jgp.55.4.497}
}

@article{sabah1969subthreshold,
  author    = {Sabah, N. H. and Leibovic, K. N.},
  title     = {Subthreshold Oscillatory Responses of the Hodgkin-Huxley Cable Model for the Squid Giant Axon},
  journal   = {Biophysical Journal},
  volume    = {9},
  number    = {10},
  pages     = {1206--1222},
  year      = {1969},
  publisher = {Elsevier},
  doi       = {10.1016/S0006-3495(69)86446-5}
}

@article{chandler1962theoretical,
  author    = {Chandler, W. K. and FitzHugh, R. and Cole, K. S.},
  title     = {Theoretical stability properties of a space-clamped axon},
  journal   = {Biophysical Journal},
  volume    = {2},
  number    = {2},
  pages     = {105--127},
  year      = {1962},
  publisher = {Elsevier},
  doi       = {10.1016/S0006-3495(62)86844-1}
}

@book{koch1999biophysics,
  author    = {Koch, Christof},
  title     = {Biophysics of Computation: Information Processing in Single Neurons},
  series    = {Computational Neuroscience Series},
  publisher = {Oxford University Press},
  address   = {New York, NY},
  year      = {1999},
  isbn      = {978-0195104912},
  doi       = {10.1093/oso/9780195104912.001.0001}
}

@book{abbot2001theoretical,
  author    = {Dayan, Peter and Abbott, Laurence F.},
  title     = {Theoretical Neuroscience: Computational and Mathematical Modeling of Neural Systems},
  series    = {Computational Neuroscience Series},
  publisher = {The MIT Press},
  address   = {Cambridge, MA},
  year      = {2001},
  isbn      = {978-0262541855}
}

@article{hodgkin1952quantitative,
  author    = {Hodgkin, A. L. and Huxley, A. F.},
  title     = {A quantitative description of membrane current and its application to conduction and excitation in nerve},
  journal   = {The Journal of Physiology},
  volume    = {117},
  number    = {4},
  pages     = {500--544},
  year      = {1952},
  publisher = {Wiley Online Library},
  doi       = {10.1113/jphysiol.1952.sp004764}
}

@book{Keener2009I,
	author = {Keener J., Sneyd J.},
	edition = {2ed},
	publisher = {Springer},
	series = {Interdisciplinary Applied Mathematics},
	title = {Mathematical physiology I: Cellular physiology},
	year = {2009}}

@book{izhikevich2007dynamical,
  author    = {Izhikevich, Eugene M.},
  title     = {Dynamical Systems in Neuroscience: The Geometry of Excitability and Bursting},
  publisher = {MIT Press},
  address   = {Cambridge, MA},
  year      = {2007},
  isbn      = {9780262090438}
}

@article{connor1977neural,
  author    = {Connor, John A. and Walter, David and McKown, Russell},
  title     = {Neural repetitive firing: modifications of the {Hodgkin-Huxley} axon suggested by experimental results from crustacean axons},
  journal   = {Biophysical Journal},
  volume    = {18},
  number    = {1},
  pages     = {81--102},
  year      = {1977},
  publisher = {Elsevier},
  doi       = {10.1016/S0006-3495(77)85598-7}
}

@article{connor1971voltage,
  author    = {Connor, J. A. and Stevens, C. F.},
  title     = {Voltage clamp studies of a transient outward membrane current in gastropod neural somata},
  journal   = {The Journal of Physiology},
  volume    = {213},
  number    = {1},
  pages     = {21--30},
  year      = {1971},
  month     = {feb},
  publisher = {Wiley},
  doi       = {10.1113/jphysiol.1971.sp009365}
}

@article{Willems1972Part2,
  author  = {Willems, Jan C.},
  title   = {Dissipative dynamical systems Part II: Linear systems with quadratic supply rates},
  journal = {Archive for Rational Mechanics and Analysis},
  year    = {1972},
  volume  = {45},
  number  = {5},
  pages   = {352--393},
  doi     = {10.1007/BF00276494}
}

@incollection{prescott2014excitability,
  author    = {Prescott, Steven A.},
  title     = {Excitability: Types I, II, and III},
  booktitle = {Encyclopedia of Computational Neuroscience},
  editor    = {Jaeger, Dieter and Jung, Ranu},
  publisher = {Springer Science+Business Media},
  address   = {New York, NY},
  year      = {2014},
  doi       = {10.1007/978-1-4614-7320-6_151-1}
}

\end{document}